\documentclass[conference]{IEEEtran}
\IEEEoverridecommandlockouts
\usepackage[utf8]{inputenc}
\usepackage{cite}
\usepackage{booktabs}
\usepackage{microtype}

\begin{document}
\title{Auditing LLM Pseudo-Label Distillation for Review Informativeness at E-Commerce Scale}

\author{
\IEEEauthorblockN{M. Emin Dogan}
\IEEEauthorblockA{\textit{Analytics Department} \\
\textit{Hepsiburada}, İstanbul, Türkiye \\
emin.dogan@hepsiburada.com}
\and
\IEEEauthorblockN{Cansu Er\textsuperscript{*}}
\IEEEauthorblockA{\textit{Analytics Department} \\
\textit{Hepsiburada}, İstanbul, Türkiye \\
cansuer.contact@gmail.com}
\and
\IEEEauthorblockN{Aysenur Kulunk}
\IEEEauthorblockA{\textit{Analytics Department} \\
\textit{Hepsiburada}, İstanbul, Türkiye \\
aysenur.kulunk@hepsiburada.com}
\and
\IEEEauthorblockN{M. Furkan Eseoglu}
\IEEEauthorblockA{\textit{Analytics Department} \\
\textit{Hepsiburada}, İstanbul, Türkiye \\
furkan.eseoglu@hepsiburada.com}
\and
\IEEEauthorblockN{H. Bahadir Sahin}
\IEEEauthorblockA{\textit{Analytics Department} \\
\textit{Hepsiburada}, İstanbul, Türkiye \\
hikmet.sahin@hepsiburada.com}
}
\maketitle
\begingroup\renewcommand\thefootnote{*}
\footnotetext{Work done while the author was at Hepsiburada.}
\endgroup

\begin{abstract}
We study hard-label distillation from Gemini 3.1 Flash-Lite to compact encoders on 5.34 million Turkish marketplace reviews. A strict new-SKU temporal test measures teacher fidelity; a separate, teacher-label-blind audit by one company annotator measures editorial agreement. An H3 category prefix slightly improves teacher fidelity on the full new-SKU test. On the 16,000-review chronological audit subset, it instead reduces agreement with both Gemini and the annotator, with a clearer paired difference for CORN than for cross entropy. Thus the context effect depends on the evaluation population, and teacher fidelity alone does not establish editorial alignment. Under the measured offline configurations, a single-L4 student pipeline processes reviews 50.2 times faster than the teacher pipeline.
\end{abstract}

\section{Introduction}
An informative review offers product-specific evidence useful to a purchase decision. A delivery-only comment gives little evidence about the product, whereas a concrete fit, durability, or compatibility observation may be informative regardless of sentiment. This construct is narrower than general helpfulness \cite{sun2019}. At marketplace scale, language models can label millions of reviews for smaller models, but a student's agreement with its teacher and its agreement with an editorial reference answer different questions.

We examine these questions using Turkish reviews, frozen Gemini pseudo-labels, a SKU-disjoint temporal test, and a separate single-annotator audit. The contribution is an evaluation of what survives entity and time shift, how input context affects the result, and the throughput of the resulting offline scoring path. A combined market/language shift and a teacher-relative ranking benchmark provide further operational checks. The audit lets us test whether a gain on the full test population holds on annotated reviews.

\section{Related Work}
Weak supervision creates training data from imperfect signals \cite{ratner2020}; hard-label LLM-to-encoder training is a practical form of distillation \cite{hinton2015}. We compare BERTurk \cite{schweter2020}, multilingual XLM-R \cite{conneau2020}, and Turkish encoders with cross entropy (CE) and conditional ordinal regression (CORN) \cite{shi2023}. Our emphasis is evaluation against distinct targets rather than a new model architecture.

\section{Data and Protocol}
The source contains 5,344,302 reviews dated June 2025--June 2026. A cutoff of May 7, 2026 was fixed before training. Historical SKUs were assigned exclusively to train (3,844,147 reviews), validation (486,000), or IID test (479,724); all six pairwise SKU intersections were checked to be empty. The strict out-of-time (OOT) set retains 159,570 reviews from SKUs first seen after the cutoff. Excluding 374,861 future reviews of previously seen SKUs leaves 4,969,441 reviews in this protocol. The 37-day strict OOT window contains 139,258 SKUs, only 1.15 reviews per SKU on average; it primarily tests classification for newly appearing products. Variants with distinct catalog SKUs remain distinct entities.

The frozen Vertex AI \texttt{gemini-3.1-flash-lite} teacher received review text, product name, and product type, but not rating, media, votes, freshness, or the production score. The prompt defines three ordinal classes: 0, no concrete product evidence; 1, a limited relevant attribute; and 2, specific actionable evidence. Temperature and thinking budget were both set to zero. Structured output requires \texttt{informativeness\_score}, \texttt{reason}, and \texttt{signals}; the nine allowed signal types cover use, measurement, attributes, comparison, pros/cons, product terms, emotion only, shipping only, and short text. Only hard scores train the students. Model ID, prompt revision, and generated labels were frozen before training; reasons and signals were never student inputs or targets.

One native-Turkish company annotator, selected for domain expertise, labeled 40,000 reviews spanning 1,189 product types without seeing Gemini's labels. No calibrated pilot preceded the audit. The annotator saw the same text and product metadata as Gemini; there was no second annotator or adjudication. The audit contains 4,000, 12,000, 8,000, and 16,000 reviews from the original train, validation, IID, and chronological OOT pools. The 16,000 future reviews were not used for training or checkpoint selection. They are temporally held out, but only 4,748 belong to the strict new-SKU set; 11,252 concern previously observed SKUs.

BERTurk students use review text alone (T) or an H3 category prefix plus text (C). CE and CORN are crossed with both inputs over seeds 17, 42, and 73. XLM-R, COSMOS Turkish BERT, and Turkish ModernBERT use CORN+C, giving 21 runs (see Tables I and III). All use 64-token truncation, AdamW at $2\times10^{-5}$, batch size 64, and at most three epochs; validation QWK against Gemini selects checkpoints. We report accuracy, macro-F1, mean absolute ordinal error (MAE), and quadratic weighted kappa (QWK). For audit contrasts, we resample SKUs 1,000 times, keeping the same draws for paired models and averaging differences across the three fixed seeds. The resulting percentile intervals reflect audit-sample uncertainty conditional on those seeds. CORN estimates $P(y>0)$ and $P(y>1\mid y>0)$; ranking uses the expected class, rescaled to 0--100. Run records retain model revisions, seeds, split assignments, checkpoints, hardware, and per-review class probabilities.

Table~\ref{tab:strict} measures reproduction of Gemini labels, not human accuracy. The category prefix raises BERTurk QWK by .0023 with CE and .0016 with CORN on the full strict OOT set. ModernBERT+CORN+C has the highest teacher fidelity. These small gains are descriptive across three training seeds. Paired category-minus-text QWK changes are $.00226\pm.00065$ for CE and $.00156\pm.00056$ for CORN; CORN offers no practically meaningful advantage over matched CE here. The pooled ModernBERT Class-2 F1 is .8681, below its Class-0 F1 of .9626.

\section{Results}
\begin{table}[t]
\caption{Teacher fidelity on 159,570 strict new-SKU OOT reviews. Three-seed means; T=text, C=category prefix.}
\label{tab:strict}
\centering\footnotesize
\setlength{\tabcolsep}{3pt}
\begin{tabular}{@{}lrrr@{}}
\toprule
Student & QWK & Acc. (\%) & F1 (\%)\\
\midrule
BERTurk CE/T & .9175 & 92.32 & 89.90\\
BERTurk CE/C & .9198 & 92.53 & 90.13\\
BERTurk CORN/T & .9173 & 92.26 & 89.87\\
BERTurk CORN/C & .9189 & 92.41 & 90.01\\
ModernBERT CORN/C & \textbf{.9223} & \textbf{92.71} & \textbf{90.42}\\
\bottomrule
\end{tabular}
\end{table}

A separate full chronological OOT analysis groups reviews into 33 buying categories. Category-aware BERTurk reaches category-macro QWK .9110, with .8626 in its worst category; these are teacher-relative results on a different population from Table~\ref{tab:strict}.

\begin{table}[t]
\caption{Agreement with one annotator on 16,000 chronological OOT reviews. Student values are three-seed means.}
\label{tab:audit}
\centering\footnotesize
\setlength{\tabcolsep}{2.5pt}
\begin{tabular}{@{}lrrrr@{}}
\toprule
System & QWK & Acc. (\%) & F1 (\%) & MAE\\
\midrule
Gemini & .9324 & 93.79 & 91.67 & .0623\\
BERTurk CE/T & .8513 & 87.17 & 82.69 & .1311\\
BERTurk CE/C & .8502 & 87.16 & 82.69 & .1312\\
BERTurk CORN/T & \textbf{.8530} & 87.14 & \textbf{82.72} & \textbf{.1310}\\
BERTurk CORN/C & .8490 & 86.85 & 82.40 & .1342\\
\bottomrule
\end{tabular}
\end{table}

Gemini agrees exactly with the annotator on 90.34\% of all 40,000 audited reviews and 93.79\% of the chronological OOT subset. The pooled 40K rate reflects fixed original-split quotas, not a random sample of all traffic. Of the 3,864 full-audit disagreements, 3,855 are between adjacent classes. Table~\ref{tab:audit} provides all four metrics on the held-out audit subset. For CE, adding category changes annotator QWK by $-.0011$ (paired bootstrap 95\% CI $[-.0039,.0012]$) and accuracy by $-.01$ percentage points. For CORN, the changes are $-.0040$ ($[-.0068,-.0014]$) and $-.29$ percentage points. These intervals are unadjusted and do not measure variation from retraining with new seeds.

On these same 16,000 reviews, category also \emph{reduces teacher fidelity}: QWK changes by $-.0011$ for CE (paired bootstrap 95\% CI $[-.0035,.0014]$) and $-.0044$ for CORN ($[-.0069,-.0019]$). On the 4,748-review audited new-SKU subset, CORN category-minus-text QWK changes are $-.0080$ against Gemini and $-.0076$ against the annotator; CE also declines on both targets. Thus the full strict-set gain and the audit-set decline describe different populations; this experiment does not show a same-population trade-off between teacher fidelity and editorial agreement.

On the 16K audit, BERTurk CORN/T has Class-0/1/2 F1 of .933/.792/.757, so Class 2 remains the hardest boundary. Across the full audit, Gemini labels 2,207 disagreements below and 1,657 above the annotator. Teacher-lower cases frequently carry shipping-only, emotion-only, or too-short signals; teacher-higher cases often carry product-attribute or use-experience signals. These teacher-generated tags help diagnose differences in the rubric but do not adjudicate which label is correct.

\begin{table}[htbp]
\caption{Teacher-fidelity QWK under a combined external shift; C=category, T=text. Amazon has no human labels.}
\label{tab:amazon}
\centering
\footnotesize
\setlength{\tabcolsep}{3pt}
\begin{tabular}{@{}lrr@{}}
\toprule
CORN student & Turkish OOT & Amazon\\
\midrule
ModernBERT/C & \textbf{.9223} & .6871\\
XLM-R/C & .9145 & \textbf{.7614}\\
COSMOS BERT/C & .9000 & .6411\\
BERTurk/T & .9173 & .5974\\
BERTurk/C & .9189 & .5552\\
\bottomrule
\end{tabular}
\end{table}

\subsection{Combined Market and Language Shift}
The breadth-oriented Amazon sample covers eight domains and 5,377 leaf product types; two failed teacher labels leave 29,998 reviews \cite{hou2024}. With seed 42, sampling first covers each observed leaf type and then fills remaining positions from capped per-type reservoirs; this is not a representative Amazon sample. Table~\ref{tab:amazon} shows that XLM-R has the highest teacher fidelity on this sample, whereas ModernBERT leads on Turkish new-SKU OOT. Class 2 accounts for 35.22\% of Amazon labels versus 11.28\% of Turkish OOT labels. XLM-R reaches 75.35\% accuracy and 75.78\% macro-F1 against those external teacher labels. Language, marketplace, taxonomy, vocabulary, and class prior all change together, so the table cannot isolate a language effect or establish external human agreement.

\subsection{Teacher-Relative Ranking}
Table~\ref{tab:ranking} uses 534,430 chronological future reviews and Gemini pseudo-labels as relevance judgments \cite{jarvelin2002}. NDCG is macro-averaged over SKUs with at least $k$ reviews and a nonzero label: 13,565, 3,274, and 736 SKUs qualify at $k=5,10,20$. This benchmark includes previously seen SKUs and is separate from the strict new-SKU test. BERT-Join is a separate BERTurk model with a CORN head that combines the review-text representation with 16 engineered numeric features, including length, measurement, and product-related signals. It ranks reviews by the expected ordinal score; its inputs and training setup differ from the 21 text/H3 runs reported above.

The semantic scores order Gemini labels better than the deployed production composite on this artifact. Students were trained on Gemini-derived labels, the production score also includes a separate Gemini richness feature, and the systems use different inputs. Thus the high NDCG values show teacher-relative ordering, not independent ranking quality; low within-SKU label diversity can also inflate them. At $k=10$, Class-2 precision is 19.73\% for BERTurk CORN/C versus 17.57\% for the production composite, a smaller gain than the graded NDCG difference. Of 324,795 SKUs in this artifact, 76.01\% have a single review and cannot enter these NDCG groups. The production composite uses media, votes, freshness, length, and a separate Gemini richness signal.

\begin{table}[htbp]
\caption{Teacher-relative macro NDCG on eligible SKU groups.}
\label{tab:ranking}
\centering\footnotesize
\setlength{\tabcolsep}{3pt}
\begin{tabular}{@{}lrrr@{}}
\toprule
Ranker & @5 & @10 & @20\\
\midrule
Random & .5499 & .5109 & .4881\\
Gemini richness & .8830 & .8702 & .8462\\
Production composite & .8095 & .7858 & .7694\\
BERT-Join & \textbf{.9880} & .9844 & .9795\\
BERTurk CORN/C & .9877 & \textbf{.9846} & \textbf{.9809}\\
\bottomrule
\end{tabular}
\end{table}

\section{Operational Scope and Limitations}
The teacher measurement covers 858,904 reviews newly scored in a resumable labeling run: rows with existing score or attempt outputs were skipped. With up to 64 concurrent API calls, the run took 5.3986 hours (44.19 newly scored reviews/s, including resume-scan overhead). A BERTurk student processed 100,000 reviews at 2,218 reviews/s on one L4 GPU, including tokenization and data transfer but excluding one-time loading. The observed pipeline-rate ratio is 50.2$\times$ under these configurations; it is neither hardware normalized nor an online latency or production A/B result. Recorded single-L4 training lasted 7.0--11.1 hours per run across the compared encoders. At batch size 128, the student's 10K-review rate was 2,247 reviews/s, only 1.3\% above its 100K rate; peak allocated GPU memory was 795 MB.

Editorial agreement is with one teacher-label-blind annotator, not an estimate of inter-annotator reliability. The annotator and teacher saw product name and type, whereas students saw only text or text with an H3 prefix. Given this input mismatch, our finding is a partial-context comparison; divergence may partly stem from information asymmetry. The 16K audit is chronologically held out, while the full strict OOT is SKU-disjoint; neither comparison should silently stand in for the other. The external Amazon test combines several shifts and cannot isolate a language effect.

\section{Conclusion}
Compact encoders reproduce frozen LLM labels on new SKUs, with measured offline throughput 50.2 times that of the teacher. The H3 category prefix gives small teacher-fidelity gains on the full strict new-SKU test, while it lowers both teacher and annotator agreement on the chronological audit. Because those populations differ, the results do not establish a same-population trade-off between teacher fidelity and editorial alignment. The Amazon and ranking checks are also teacher-relative. Broader editorial claims require multiple annotators and matched teacher/student inputs, alongside explicit reporting of the evaluation population and reference target.

\noindent\textit{Artifact availability.}  Due to proprietary company constraints, the full teacher prompt, output schema, and experiment configurations are archived in an internal GitLab repository and are not publicly released.

\bibliographystyle{IEEEtran}
\bibliography{references}

@article{sun2019,
  title={Helpfulness of online reviews: Examining review informativeness and classification thresholds by search products and experience products},
  author={Sun, Xinyu and Han, Maoxin and Feng, Juan},
  journal={Decision Support Systems},
  volume={124},
  pages={113099},
  year={2019},
  publisher={Elsevier}
}

@article{ratner2020,
  title={Snorkel: rapid training data creation with weak supervision: A. Ratner et al.},
  author={Ratner, Alexander and Bach, Stephen H and Ehrenberg, Henry and Fries, Jason and Wu, Sen and R{\'e}, Christopher},
  journal={The VLDB Journal},
  volume={29},
  number={2},
  pages={709--730},
  year={2020},
  publisher={Springer}
}

@misc{hinton2015,
      title={Distilling the Knowledge in a Neural Network}, 
      author={Geoffrey Hinton and Oriol Vinyals and Jeff Dean},
      year={2015},
      eprint={1503.02531},
      archivePrefix={arXiv},
      primaryClass={stat.ML},
      url={https://arxiv.org/abs/1503.02531}, 
}

@misc{schweter2020,
  author       = {Stefan Schweter},
  title        = {BERTurk - BERT models for Turkish},
  month        = apr,
  year         = 2020,
  publisher    = {Zenodo},
  doi          = {10.5281/zenodo.3770924},
  url          = {https://doi.org/10.5281/zenodo.3770924},
  note         = {Software, Version 1.0.0}
}

@inproceedings{conneau2020,
  title={Unsupervised cross-lingual representation learning at scale},
  author={Conneau, Alexis and Khandelwal, Kartikay and Goyal, Naman and Chaudhary, Vishrav and Wenzek, Guillaume and Guzm{\'a}n, Francisco and Grave, Edouard and Ott, Myle and Zettlemoyer, Luke and Stoyanov, Veselin},
  booktitle={Proceedings of the 58th annual meeting of the association for computational linguistics},
  pages={8440--8451},
  year={2020}
}

@article{shi2023,
  title={Deep neural networks for rank-consistent ordinal regression based on conditional probabilities},
  author={Shi, Xintong and Cao, Wenzhi and Raschka, Sebastian},
  journal={Pattern Analysis and Applications},
  volume={26},
  number={3},
  pages={941--955},
  year={2023},
  publisher={Springer}
}

@article{jarvelin2002,
  title={Cumulated gain-based evaluation of IR techniques},
  author={J{\"a}rvelin, Kalervo and Kek{\"a}l{\"a}inen, Jaana},
  journal={ACM Transactions on Information Systems (TOIS)},
  volume={20},
  number={4},
  pages={422--446},
  year={2002},
  publisher={ACM, New York, NY, USA}
}

@misc{hou2024,
      title={Bridging Language and Items for Retrieval and Recommendation: Benchmarking LLMs as Semantic Encoders}, 
      author={Yupeng Hou and Jiacheng Li and Xiangjun Fu and Zhankui He and An Yan and Xiusi Chen and Julian McAuley},
      year={2026},
      eprint={2403.03952},
      archivePrefix={arXiv},
      primaryClass={cs.IR},
      url={https://arxiv.org/abs/2403.03952}, 
}

\end{document}